      [1999/12/01 v1.4c Il Nuovo Cimento]
\documentclass{cimento}

\usepackage{graphicx}  

\title{Generation of Medical X-ray and Terahertz Beams of Radiation
Using Table-Top Accelerators}
\author{V.~Baryshevsky\from{ins:x}\thanks{v$\_$baryshevsky@yahoo.com;bar@inp.minsk.by},
A.~Gurinovich\from{ins:x} , E.~Gurnevich\from{ins:x} ,
A.~Lobko\from{ins:x}} \instlist{\inst{ins:x} Research Institute
for Nuclear Problems, Minsk, Belarus}
\PACSes{\PACSit{41.60.-m,41.75.-i} \PACSit{42.25.-p}}

\begin{document}

\maketitle

\begin{abstract}
Theoretical and experimental studies of PXR and diffracted
radiation of an oscillator in crystals combined with the
development of VFEL generators with photonic crystals give a
promising basis for creation of X-ray and THz sources using the
same table-top accelerator.
Multi-modal medical facility can be developed on the basis of one
dedicated table-top electron accelerator of some tens of MeV
energy.
Such a system could find a lot of applications in medical practice
and biomedical investigations.

\end{abstract}


In the Executive Summary of the first Workshop ``Physics for
Health in Europe'' held in February 2010 at CERN is stressed that
dose reduction during diagnostic radiology and CT examinations is
a hot research topic worldwide \cite{bar_1}. Development of new
intensive (quasi)-monochromatic tunable x-ray and terahertz
sources is an important part of such research.

%
Medical quasi-monochromatic X-ray beams must have an essential
integral flux to provide high-quality high-contrast imaging. Also,
realistic sources of these beams must have laboratory sizes and
affordable price to be used in clinics and hospitals.
The main problem in development of monochromatic X-ray sources is
the gap between the achievable photon generation efficiency
(photon per electron) and the existing electron beam current in
the table-top accelerators.
Another problem is strong scattering of an electron moving
through a single crystal target.
We showed \cite{bar_2} that an X-ray source with the required
properties can be developed using
%
parametric X-rays (PXR) from charged particles in a crystal
%
%
\cite{bar_3} and table-top accelerators like compact storage ring
or pulsed race-track microtron \cite{bar_9,L_1,L_2,L_3,L_4}.

Accelerators of a kind can also be  used for the development of
the intense terahertz source based on the mechanism of the volume
free electron laser \cite{bar_8}.
%
Today, THz technology is finding a variety of applications:
information and communications technology; biology and medical
sciences; non-destructive evaluation and homeland security;
quality control of food and agricultural products; global
environmental monitoring, space research and ultra-fast computing
\cite{THZ-review}.
High-power tunable T-ray sources are very important devices to
bring THz research from promising prospects to a wide use in
science and technology.

 T-rays are also very promising for biomedical applications: low
energy of photons prevents them from ionizing biological media,
but this energy corresponds to vibrational levels of important bio
molecules including DNA and RNA. This allows direct action for
stimuling viruses, cells, their components and provides control of
bio-chemical reactions. Thus, T-rays may be applied in therapy,
surgery, imaging, and tomography. Terahertz radiation is extremely
important for  bio-medical applications and its wider use depends
on the progress in development of THz sources \cite{bar_10}.


In present paper we discuss prospects of application of
%
%
%
diffracted radiation of an oscillator (DRO) (sometimes called
diffracted channeling radiation - DCR)
%
%
%
for X-ray source creation.
DRO is the coherent process of diffracted X-ray photon emission by
the relativistic oscillator (electron channeling in a crystal).
DRO formation was first considered in \cite{bar_6},  detailed
review and references may be found in \cite{bar_7,lanl2010}.
Evaluations show that DRO generation efficiency per one electron
is some times higher than that of PXR and diffracted
\textit{Bremsstrahlung}.

The same radiation mechanisms can work in the terahertz range if a
single crystal target is changed for the photonic crystal of
appropriate properties
\cite{RC2005-NIM,new_paper,FEL_2002THz,bar_19}. Thus, multi-modal
medical facility can be developed on the basis of one dedicated
table-top electron accelerator of several tens of MeV energy.

\section{Generation of Medical X-ray}

The principal question arises: what do we need for high-quality in
vivo medical imaging?

Current understanding shows, that we need approximately $\sim
10^{12}$ photons/s with tunable X-ray energy in 10-70 keV spectral
range \cite{bar_1,bar_2}.
Monochromaticity should be of $\sim 10^{-2}$ and lower for a
patient's dose reduction. Radiation background should be low.

%
%
%
%

Besides the above mentioned PXR \cite{bar_2},
electrons (positrons) moving in a crystal radiate X-rays due to
two more mechanisms:

a). surface parametric x-ray radiation (SPXR)
\cite{bar_7,lanl2010,bar_5};

b). diffracted radiation of a relativistic oscillator (DRO), which
is also often called diffraction channelling radiation (DCR)
\cite{bar_6} (detailed review and references can be found in
\cite{bar_7,lanl2010}).

According to the analysis \cite{bar_7}, the DCR to PXR ratio for
electrons of 34 MeV energy channelled in a $Si$ target between
(100) planes and radiation diffracted by (220) planes is estimated
as R$\sim $5 ($R = \frac{{I_{DCR}} }{{I_{PXR}} }$).

For qualitative analysis, the DCR intensity can be evaluated as a
product of intensity of channelled radiation and reflection
coefficient for Bragg diffraction in the defined range of angles
and frequencies.

At first glance the idea for a new type of X-ray source could be
to combine two crystals: X-rays emitted from the particles
channelled in the first crystal obtain the required spectral and
angular distributions due to diffraction in the second one.

Nevertheless, the possibility of anomalous transmission of X-rays
diffracted in a crystal (Bormann effect \cite{Bormann}) should be
remembered.
In this case the coefficient of X-rays absorption becomes lower
and the intensity of DCR (DRO) can appear higher than that of
conventional channelling radiation.
The radiation intensity is proportional to the path length of the
particle inside the crystal indeed while it becomes comparable
with the absorption length (if the dechannelling length exceeds
the absorption path). The absorption path in the presence of
diffraction can be much longer than that in the absence of
diffraction. The above is especially important in the X-ray
frequency range, for which absorption in the absence of
diffraction is high.

That is why the X-ray source based on DCR mechanism could be more
powerful.

\section{Terahertz radiation (T-rays)}

The above described mechanisms of photon radiation by a
relativistic particle passing through a crystal (PXR,SPXR,DRO)
have general nature and holds for a particle moving either through
an artificial (photonic) crystal or along its surface.
Photonic crystals (diffraction gratings) can be made with various
spatial periods and thereby provide radiation in different
wavelength ranges from microwave to optical and even soft X-rays.
Note that radiation from a relativistic particle moving in either
natural or artificial crystal is  spontaneous radiation.
Obviously, the induced radiation could also exist.

Generation of radiation in millimeter and far-infrared range
with nonrelativistic and low-relativistic electron beams is a complicated task. 
Gyrotrons and cyclotron resonance facilities are used as sources
in millimeter and sub-millimeter range, but for their operation a
magnetic field of several tens of kiloGauss is necessary.
Slow-wave devices (TWT, BWT, orotrons) in this range require
application of dense and thin ($<0.1$mm) electron beams because
only electrons passing near the slowing structure at a distance $d
\leq \lambda \beta \gamma /(4\pi )$ can effectively interact with
electromagnetic waves ($\gamma$ is the Lorentz factor, $\lambda$
is the wavelength, $\beta=\frac{v}{c}$, $v$ is the particle
velocity).
For 1 THz frequency $\lambda = 3 \cdot 10^{-2}$ cm, therefore, for
an electron with energy of 50 MeV ($\gamma=10^2$), the above
discussed distance $d\leq 0.3$ cm.
It is difficult to guide thin beams near a slowing structure with
desired accuracy.
Conventional waveguide systems are essentially restricted by the
requirement for transverse dimensions of a resonator, which should
not significantly exceed the radiation wavelength. Otherwise, the
generation efficiency decreases abruptly due to the excitation of
plenty of modes. Most of the above problems can be overcome in
Volume Free Electron Lasers (VFEL)
\cite{bar_7,lanl2010,RC2005-NIM,bar_19,bar_5}.

 In  volume FELs, the greater part of the electron beam interacts with an electromagnetic wave due to
 volume distributed interaction.
Transverse dimensions of a VFEL resonator could significantly
exceed radiation wavelength $D \gg \lambda $. Multi-wave Bragg
dynamical diffraction provides mode discrimination in VFELs.

Moreover, in VFEL a lasing regime discovered in \cite{bar_11} can
be realized. It provides drastic increase in the efficiency of the
electron beam interaction with the emitted electromagnetic wave
due to the interception of roots of the dispersion equation
describing radiative instability of the electron beam in a
3-dimensional spatially periodic medium.

\begin{figure}[tbp]
\includegraphics*[width=150mm]{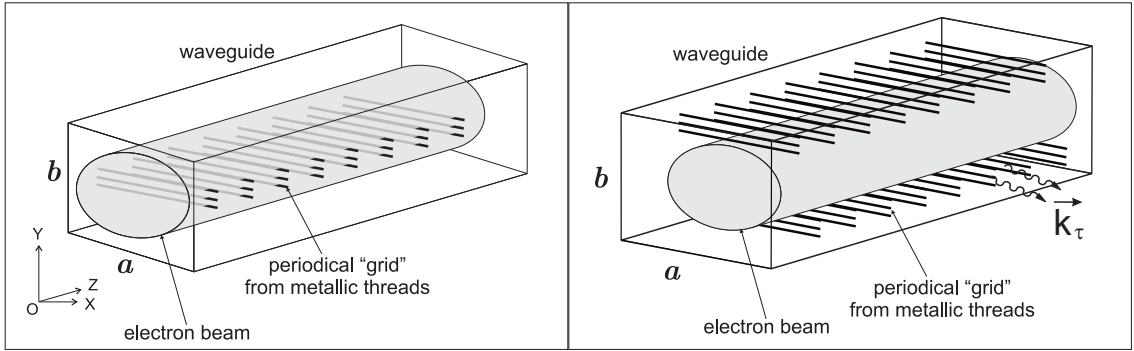}
\caption{General view of Volume Free Electron Laser with the
photonic crystal made of metal threads: a). electron beam passes
through the crystal, b). electron beam passes along the surface of
the crystal (or in the slit between two crystals)}
\label{nim03_f1}
\end{figure}

Let us consider an electron beam with velocity $\vec{u}$ passing
through a periodic structure composed of either dielectric or
metal threads (see Figure \ref{nim03_f1}).
Fields, appearing while an electron beam passes through a volume
spatially periodic medium, are described by the set of equations
given in \cite{bar_7,lanl2010,laser_8}. Instability of the
electron beam is described by the dispersion equation
\cite{bar_7,lanl2010,new_paper,bar_12,bar_13,laser_8}:
\begin{equation}
(k^{2}c^{2}-\omega ^{2}\varepsilon )(k_{\tau }^{2}c^{2}-\omega
^{2}\varepsilon +\chi _{\tau }^{(b)})-\omega ^{2}\chi _{\tau }\chi
_{-\tau }=0,  \label{nim03_disp}
\end{equation}
$\vec{k}_{\tau }=\vec{k}+\vec{\tau}$ is the wave vector of the
diffracted photon, $\vec{\tau}=\left\{\frac{2\pi }{a}l;\frac{2\pi
}{b}m;\frac{2\pi }{c}n\right\}$
 are the reciprocal lattice vectors,
$a,b, c$ are the translation periods,
 $\chi _{\alpha }^{(b)}$ is the part of dielectric
susceptibility caused by the presence of the electron beam.
%
%
Two different types of instability exist, depending on the
radiation frequency. Amplification takes place when the electron
beam is in synchronism with the electromagnetic component
$\vec{k}+\vec{\tau}$, which has a positive projection $k_{z}$. If
the projection $k_{z}$ is negative and the generation threshold is
reached, then generation evolves. In the first case, radiation
propagates along the transmitted wave which has a positive
projection of group velocity
$v_{z}=\frac{c^{2}k_{z}^{(0)}%
}{\omega }\quad (k_{z}^{(0)}=\sqrt{\omega ^{2}\varepsilon
-k_{\perp }^{2}})$,
 and beam disturbance moves along it. In the second case,
the group velocity has a negative projection
$v_{z}=-\frac{c^{2}k_{z}^{(0)}}{\omega }$,
 and radiation propagates
along the back-wave and the electromagnetic wave comes from the
range of the greatest beam disturbance to the place, where
electrons come into the interaction area. For a one-dimensional
structure, such a mechanism is realized in a backward-wave tube.
In amplification case, equation (\ref{nim03_disp}) gives for the
increment of instability: $\texttt{Im} k_{z}^{\prime
}=-\frac{\sqrt{3}}{2}f $, where
\[
f=\sqrt[3]{\frac{h\omega _{L}^{2}(\vec{u}\vec{e}%
^{\tau })^{2}\omega ^{4}r}{2k_{z}^{(0)}c^{4}u_{z}^{2}\left(
k_{\tau }^{2}c^{2}-\omega ^{2}\varepsilon _{0}\right) }},
\]
 if the condition
$ 2k_{z}^{\prime }f\gg \frac{\omega ^{2}\chi _{0}"}{c^{2}} $ is
fulfilled. Here $r=\chi_{\tau}\chi_{-\tau}$,
$h(\vec{u}\vec{e}^{\tau})^2/c^2=1/\gamma^3$ if the electron beam
propagates in a strong guiding magnetic field, otherwise,
$h=1/\gamma$, $\vec{u}$ is the beam velocity. In case $
2k_{z}^{\prime }f\ll \frac{\omega ^{2}\chi _{0}"}{c^{2}}$ a
dissipative instability evolves. Its increment is
\[
\texttt{Im}
k_{z}=-\frac{c}{\omega}\sqrt{\frac{k_{z}^{(0)}f^{3}}{\chi_{0}}}.
\]
If inequalities $k_{z}^{`2}\gg 2k_{z}k_{z}^{`}$ and $k_{z}^{`2}\gg \frac{%
\omega ^{2}\chi _{0}"}{c^{2}}$ are fulfilled, the spatial
increment of instability can be expressed as
\[ \texttt{Im}
k_{z}^{`}=-\left( \frac{h\omega _{L}^{2}(\vec{u}\vec{e}^{\tau
})^{2}\omega ^{4}r}{c^{4}\left( k_{\tau }^{2}c^{2}-\omega
^{2}\varepsilon _{0}\right) u_{z}^{2}}\right) ^{1/4},
\]
but the parameters providing such dependence correspond to the
conversion from the amplification to the generation regime (for
the Compton instability
this situation takes place at $%
k_{z}^{(0)}\approx 0$). The frequency of amplified radiation is
defined as:
\begin{equation}
\omega =\frac{\vec{\tau}\vec{u}}{1-\beta _{x}\eta _{x}-\beta
_{y}\eta _{y} - \beta _{z}\sqrt{\varepsilon -\eta _{x}^{2}-\eta
_{y}^{2}}}. \label{nim03_freqa}
\end{equation}
The instability in the generation regime is described by the
temporal increment and cannot be described by the spatial
increment. The increment of absolute instability can be found by
solving the equation $\texttt{Im} k_{z}^{(+)}(\omega )=\texttt{Im}
k_{z}^{(-)}(\omega )  $ with respect to the imaginary part of
$\omega $ \cite{lanl2010}.

The use of Bragg multi-wave distributed feedback increases the
generation efficiency and provides discrimination of generated
modes.  If the conditions of synchronism and Bragg conditions are
not fulfilled simultaneously, diffraction structures with
different periods can be applied \cite{lanl2010,nim03_nonrel}. One
of them provides synchronism of the electromagnetic wave with the
electron beam $\omega -\vec{k}\vec{u}=\vec{\tau}_{1}\vec{u}$. The
second diffraction structure evolves distributed Bragg
coupling $|\vec{k}%
|\approx |\vec{k}+\vec{\tau}_{j}|$, $\vec{\tau}_{j}$ ($j=2\div n$)
are the reciprocal lattice vectors of the second structure.

Under dynamical diffraction conditions either the generation start
current or the length of the generation zone at certain values of
the current can be reduced \cite{bar_7,lanl2010,bar_5,bar_11}.

Each Bragg condition holds one of free parameters. For example,
for  certain geometry and electron beam velocity, two conditions
for three-wave diffraction entirely determine transverse
components of wave vectors $k_{x}$ and $k_{y}$, and therefore the
generation frequency. Hence, volume diffraction system provides
mode discrimination due to multi-wave diffraction.

 The above results affirm that a volume diffraction structure provides
both amplification and generation regimes even in the absence of
dynamical diffraction. In the latter case, generation evolves with
backward wave, similarly to the backward-wave oscillator. The
frequency in such structures is changed smoothly either by  a
smooth variation of the radiation angle (variation of $k_x$ and
$k_y$), or by the rotation of the diffraction grating or the
electron beam. For certain geometry and reciprocal lattice vector,
amplification corresponds to higher frequencies than generation.
Rotation of either the diffraction grating or the electron beam
also changes the value of the  boundary frequency, which separates
generation and amplification ranges. The use of multi-wave
distributed feedback owing to Bragg diffraction allows one either
to increase the generation efficiency or to reduce the length of
the interaction area.

%
%
In the case of our interest, a T-ray source generating either
backward or following waves with a wavelength of 0.3 mm
should have the period of the diffraction grating providing Bragg
coupling  $\sim $ 0.16 mm  that is a challenge.

It is interesting that according to
\cite{RC2005-NIM,new_paper,bar_19}, for a photonic crystal made
from metallic threads, the coefficients $\chi \left( {\tau}
\right)$, defining the threshold current and the growth of the
beam instability, are practically independent of $\tau $ up to the
terahertz range of frequencies because the diameter of the thread
can easily be made smaller than the wavelength.
As a result, an electromagnetic wave is sufficiently strongly
coupled with the diffraction grating (the coefficient of
diffraction reflection is sufficiently high) even for $\lambda \ll
d$ ($d$ is the photonic crystal (diffraction grating) period) to
provide efficient radiation in Backward Wave Oscillator (BWO) and
Travelling Wave Tube (TWT) regimes.
That is why photonic crystals with the period of several
millimeters can
 be used for lasing in terahertz range at high harmonics (for
example,  photonic crystal with a 3 mm period provides the
frequency of the tenth harmonic of about 1 terahertz ($\lambda
$=300 micron) \cite{new_paper}.

Note here that a photonic crystal built from metallic threads has
the refractive index $n_{0} < 1$ for a wave with the electric
polarizability parallel to the threads, i.e., in this case the
Cherenkov instability of the beam does not exist \cite{RC2005-NIM}
and radiation appears only due to  diffraction of waves. But if
the electric vector of the wave is orthogonal to the metallic
threads, the refractive index is $n_{0}
> 1$ , so for such a wave, the Cherenkov instability exists
\cite{22-Zhenya} even in the absence of diffraction.

The ''grid'' structure formed by periodically strained dielectric
threads was experimentally studied in \cite{nova_bar4}, where it
was shown that ''grid'' photonic crystals have sufficiently high
$Q$ factors ($10^4 - 10^6$).
Lasing from  VFELs  with the ''grid'' resonator formed by
periodically strained  metal threads was observed in~\cite{bar_8}.

Use of the volume distributed feedback in a photonic crystal makes
available:

1. frequency tuning at fixed energy of the electron beam in a
significantly wider range than conventional systems can provide;

2. significant reduction of the threshold current of the electron
beam due to more effective interaction of the electron beam with
the electromagnetic wave,  allowing, as a result, miniaturization
of generators;

3. simultaneous generation at several frequencies;

4. effective modes selection in oversize systems, in which the
radiation wavelength is significantly smaller than the resonator
dimensions.

\end{document}